\documentclass[journal, a4paper, 12pt]{IEEEtran} 

\usepackage{amsmath} 
\usepackage{graphicx, multirow, multicol}

\usepackage{color}
 \fontfamily{cmtt} \selectfont

\def\say[#1]{\iftrue \subsubsection{#1} \fi}
\def\sep[#1]{\iffalse {\noindent {\color{blue} \tiny \texttt{#1} {\leaders\hbox{\rule{2pt}{0.4pt} } \hfill} }} \newline \indent  \fi}

\begin{document}


\title{\Large \sc Navigating dark liquidity \\ \small - How Fisher catches Poisson in the Dark - }
\author{\small \sc Ilija I. Zovko, Barclays London\\ { ilija.zovko@barclays.com}}

\maketitle

\begin{abstract}
\newline
\sep[Liquidity quality vs. quantity trade-off]
In order to reduce signalling, traders may resort to limiting access to dark venues and imposing limits on minimum fill sizes they are willing to trade. However, doing this also restricts the liquidity available to the trader since an ever increasing quantity of orders are traded by algos in clips (small sized trades). An alternative is to attempt to monitor signalling in real time and dynamically make adjustments to the dark liquidity accessed.

\sep[Pitfalls of using slippage]
In practice, price slippage against the order is commonly taken as an indication of signalling. However, estimating slippage is difficult and requires a large number of fills to reliably detect it. Ultimately, even if detected, it fails to capture an important element of causality between dark fills and lit prints - a signature of information leakage. In the extreme, this can lead to scaling back trading at a time when slippage is caused by a competing trader consuming liquidity, and the appropriate action would be to scale trading up -- not down -- in order to capture good prices. 

\sep[Aim of the methodology]
In this paper we describe a methodology aimed to address this dichotomy of trading objectives, allowing to maximally capture available liquidity while at the same time protecting the trader from excessive signalling. The method is designed to profile dark liquidity in a dynamic fashion, on a per fill basis, in contrast to historical venue analyses based on estimated slippage. This allows for a dynamic and real-time control of the desired liquidity exposure.

\sep[Principles]
The method aims to detect signalling in a more direct fashion than by estimating slippage, which is a consequence of signalling in certain circumstances only. This we propose to do by measuring the time between the dark fill and the next lit print. This duration is then contrasted with the background lit trading activity at the time in order to quantify the surprise factor of the lit trade timing. The intuition is that dark fills, quickly followed by lit prints are leaking information and likely triggering the lit trading activity -- opposite to what quality dark liquidity should be.
	
\sep[Backtest results]
An empirical validation of this approach reveals that dark fills quickly followed by lit trades -- quantified by this methodology to be an indication of signalling -- suffer about 3 times higher post fill slippage than dark fills for which there is no evidence of signalling. In terms of algo performance, in as realistic a setting as a backtest allows, the methodology results in around a 30\% improvement in average slippage value for dark only executions. The additional real benefit is also an increased access to liquidity, which is difficult to quantify in a backtest.
\end{abstract}

\section{Introduction}
\sep[Why trade in dark]
The proliferation of dark trading venues has in part been driven by the contrasting desires of traders to maximise liquidity access, while at the same time minimising information of their full trading intention leaking to other market participants. This led to the practice of profiling dark pools by the "quality" of liquidity available. The property sought is the ability to trade large quantities with minimal price impact, i.e., to match large opposing orders without visibility to the lit market, until after the deal has been done. While matching in size may still be the ideal, the proliferation of algo executions like VWAP or TWAP\footnote{Volume- and Time- Weighted Average Price} meant that the liquidity of large opposing orders may also have to be accessed in multiple small fills.

\sep[Not all liquidity is alike]
The main problem for traders is that among the small slices of VWAPs and TWAPs, information seeking liquidity may be interspersed, which often does not represent a genuine trading intent. On the contrary, by filling a small size in a dark venue, the liquidity provider learns about large orders trading and can position their own books appropriately to profit from the knowledge. This, more often then not, has an effect similar to front running and causes additional market impact. When filled against these information seeking orders, traders will systematically suffer the consequences of information leaking.

\sep[How do traders avoid signalling?]
In order to avoid or limit this kind of signalling, it is not uncommon for traders to require a minimum fill quantity to their dark orders as filling larger sizes usually proves too costly for information seeking liquidity providers. Rough estimates reveal that dark fills in sizes up to \pounds5,000 on average suffer around 1bp in 5 second post fill slippage. This however reduces to zero for fill sizes larger than \pounds30,000.\footnote{Calculations based on Barclays dark fills in 2016.}

\sep[Limiting available liquidity]
However, requiring a minimum fill size \emph{in a non-discriminatory fashion} can significantly reduce available liquidity. An opposing VWAP algo order, for example, will be sliced into multiple small trades in order to best track the benchmark. Imposing a minimum fill size will prevent the trader from accessing the liquidity offered by this algo order altogether.

\sep[Objective and challenge]
The objective, therefore, is to maximise access to and trading with genuine liquidity, while limiting interaction with information seeking liquidity. The challenge is that it is difficult to distinguish between the two types of liquidity and as a result defensively limiting one's access to liquidity.

\sep[Liquidity is a dynamic property]
Given the intricacies and complexity of today's liquidity landscape, the answer most likely isn't as simple as historical profiling and restricting access to only a small subset of venues. Neither is it to impose an overall minimum fill size based on a historical analysis. Both of these approaches are static in nature and are ill suited to dynamically react to the changing composition of liquidity.

As traders enter and exit the market, the instantaneously available liquidity on each venue changes -- it is not a static market property. The appropriate mechanism by which a dark liquidity trader seeks trading counterparts should therefore also be a dynamic process. With each fill, the trader needs to assess the type of liquidity they are interacting with and make dynamical routing decisions to maximise fill rates and minimise information leakage.

Market structure changes due to MiFID II regulation will most likely only exacerbate the problems, making it all too important to have a methodology in place to properly interact with changing liquidity -- discriminating between venues with different mechanisms and dynamically in time.

\section{Profiling via post trade slippage}
A common way of dynamically evaluating liquidity is to measure average post trade price slippage.\footnote{Interchangeably, terms like impact, slippage, decay, reversion, are all used to describe it.} This is typically done by computing the post fill signed price move at an appropriate time horizon $\tau$ (usually short), and then potentially grouping the fills by venue or size bucket, arrive to an average slippage value per group. In other words, it is the average price move, conditional to the sign and time of the fill

\begin{equation}
\mathcal{R}(\tau) = E \left[ \epsilon_t \cdot (p_{t+\tau} - p_t ) | \epsilon_t \right].
\end{equation}
Here $E[ \cdot ]$ denotes averaging over the fills at times $t$, $\tau$ is the short term horizon of interest, $p_t$ the log mid price at time $t$, and $
\epsilon_t$ the buy-sell indicator (-1 for a sell, +1 for a buy) of the fill at time $t$. 

\sep[Issue of noise when measuring slippage]
A difficulty with this approach in practice lies with the fact that fairly long fill histories need to be traded in order to reliably determine the existence or lack of slippage. This precludes the metric from being reactive -- and as a consequence not very useful in practice. If we are able to detect that the liquidity we interacted with was leaking information \emph{only after the interaction}, we cannot use this knowledge to dynamically react with the aim of reducing signalling and improving performance.

\sep[Number of fills required to detect slippage]
The reason why estimating slippage in practice requires many fills lies in the relative magnitudes of the slippage and price return variance. In order to reliably ascertain that $\mathcal{R}(\tau)$ is different from zero, implying the existence of systematic price movement following a fill, a t-test is commonly used. This entails comparing the mean price slippage post fill $\mu$ to the standard deviation of unconditional price returns $\sigma$. The fact that the mean grows linearly with the number of fills $T$, while the standard deviation as a square root, imposes a lower bound on the number of data points required to detect significance as
\begin{equation}
\frac{T \mu}{\sqrt{T} \sigma} = 1 
\end{equation}
which, solving for $T$, becomes
\begin{equation}
\sqrt{T} = \frac{1}{\mu/\sigma} 
\label{Tdependency}
\end{equation}
Interpreting this as the time required to detect a signal of slippage in a noisy market environment, the minimum time is at least equal to $(\sigma/\mu)^2$ or $1/\text{Sharpe}^2$~\cite{farmer}. The smaller the Sharpe ratio, the longer it will take to significantly detect a signal of slippage, and detecting a half as strong signal takes four times longer.

\sep[Practical example]
Making things further difficult is that this low bound assumes an unrealistic 100\% trade participation rate, with a fill at each market price change. With a more realistic participation rate things get worse. A back of an envelope calculation reveals that in typical circumstances hundreds of fills are required to statistically detect slippage. A typical order of magnitude for impact per fill is $\mu=0.5$bp (basis points, see Fig~\ref{slipVsPval}), while price return standard deviation \emph{per return} is about $\sigma_1=3$bp. Rough estimates from market data reveal that there are about 3 quote returns per trade.\footnote{These rough values are based on Lloyds Bank (LLOY) data in London on March 10th 2017.} Trading an order at 20\% participation, we would roughly expect to receive a fill for each 5 market trades, corresponding to $3 \cdot 5=15$ quote returns. The standard deviation of market returns at this time horizon is $\sigma_{15} = \sigma_1 \sqrt{15} \sim 12 \text{bp}$. Plugging these rough numbers into the above expression for the minimum number of fills required to statistically detect slippage we get
\begin{equation}
T = \frac{1}{(\mu/\sigma_{15})^2} = \left(\frac{12}{0.5}\right)^2 \sim 570 \text{ fills.}
\end{equation}

Practically speaking, \emph{a trader needs to receive hundreds or more fills before they can reliably determine that the fills are causing slippage.} Any execution with less than this number of fills does not result in a sufficient number of data points that would make it possible to reliably detect potential slippage it may be causing, and in time to make any reactive adjustments to the execution.

As a consequence, dynamic executions that monitor slippage in practice tend to react more to random price drifts than to signalling, simply because there hasn't elapsed enough time for the sufficient number of fills to be collected in order to reliably detect signalling.

\sep[Slippage is not causality]
An altogether different issue with slippage, but conceptually possibly a more important one, is that slippage may, or may not, be an indication of signalling. This is important because, if the slippage is \emph{not caused} by the dark fill signalling, it is more likely an indication of increased trading pressure from the same side as the dark order. Arguably, pulling back trading in order to reduce slippage, not only will it fail to reduce slippage (because it is not causing it), it will also fail to capture good available prices, losing out on the opportunity to trade at favourable prices. 

Misattributing observed slippage to own dark fills, when in fact slippage may, at best, be random market fluctuation, or at worst, caused by a competing trader consuming the available liquidity, may result with the trader scaling their trading back at a time when they should arguably scale it up to capture good prices in an escaping market. The consequence, of course, will be inferior average performance of dark orders.

\section{Detecting dark fill signalling}
The methodology we present here addresses the aforementioned shortcomings of using slippage to dynamically improve dark liquidity trading.

On the one hand, it attempts to directly address the question if the dark fills are \emph{causing} information leakage -- the answer to which provides the basis for a sound decision to scale trading up or down. 

On the other hand, the methodology uses the valuable data generated by each fill in a statistically efficiently manner. In this way the trader does not have to wait long to amass the evidence for leakage. Each fill, while bringing in valuable information, is at the same time potentially costly and detrimental to the overall performance. Detecting leakage with only a small set of fills makes the methodology practically relevant and cost effective.

\sep[Basic premise]
The basic presumption is that trades in the lit market, \emph{quickly following a fill in the dark} are indicative of information leakage. Intuitively speaking, by filling in a dark venue, the trader does not expect to see trade prints in the lit markets, barring from the usual background lit trading activity at the time. Each lit trade following a dark fill therefore carries with it a "surprise" factor -- or a lack of it. For example, it would be surprising to observe a lit print, say 10ms, just after we've received a dark fill if the usual trading activity at the time was one print each second on average. It would, conversely, \emph{not} be a surprise if the next lit print happens half a second after the dark fill.

\sep[Dark-lit sweep]
Extremely short time periods between the dark fill and lit print, also captured by this methodology, are also indicative of the \emph{Dark-lit sweep} orders available on many trading platforms. It is obviously not very advantageous to fill passive dark orders against dark-lit sweeps because of the immediate impact they generate.

\sep[Venue latency]
Finally, in the same spirit as the surprise with lit prints following a dark fill, lit prints immediately \emph{preceding} a dark fill can be indicative of being filled at latent prices. Again, good reason to avoid being filled on such venues, or against latency sensitive liquidity.

\sep[Direction of lit prints in relation to dark]
The direction (buy or sell) of the lit prints in relation to the dark fill, depending on the decision of the trader may, or may not, play a role.

Dark fills causing lit prints in the same direction may indicate that the dark order is being front run by the dark fill counterpart. The appropriate actions, arguably, would be to scale the trading back, increase the minimum required fill size, or move the dark trading to a different venue altogether.

If the caused trades are in the opposite direction, as would be observed from a Dark-lit sweep, arguably the trading should be scaled back as well. Dark-lit sweeps may be indicative of opposing orders and by scaling back trading and waiting, the trader may be able to capture better prices later, after the opposing orders have caused market impact. Alternatively, there may be circumstances where the trader may choose to increase the trading rate to complete the order quicker.

Whichever the case may be, having the full information of causality and the relative direction of lit and dark trades, sheds light on the dark liquidity a trader is interacting with and allows for the most appropriate trading decision to be made.

\sep[Statistical test intuitively described]

As mentioned, a dark fill immediately preceded by or quickly followed by lit prints is \emph{surprising}, with the surprise factor determined by looking at the typical lit market trade rate around the time of the dark fill in question.

Obviously, based on \emph{only one} surprising dark fill, we can not make a reliable judgment about signalling. The lit market print could have by chance occurred just after a dark fill. However, upon receiving a number of such surprising fills, we can with a degree of certainty determine that there may be information leakage caused by the fills and take appropriate action. 

The number of surprising fills required to reliably detect signalling is determined by the level of surprise each fill brings. The more surprising fill and lit trade timings are, the less fills we need to determine signalling.

\section{Fisher causality metric}
The methodology rests on the assumption that lit market prints are a Poisson point process with dynamic intensity. This means that trades arrive randomly in time with a rate $1/\lambda(t)$. This assumption can be relaxed by using Hawkes self exciting processes in place of the Poisson, but in our view, the dynamic Poisson process, together with a careful preparation of the data is quite sufficient for the purpose.\footnote{Estimating the local Poisson rate with only a small number of recent market trades, results in the estimate being very dynamic, alleviating concerns that market prints are not well represented by an i.i.d. process. Furthermore, the flip side of using a small number of data points is compensated by the repeated nature of the experiment corresponding to the multiple dark fills.}

\sep[Poisson process and the Waiting paradox]
For a Poisson process the time \emph{durations} between trades are distributed as an exponential density \mbox{$1/\lambda \cdot e^{-t/\lambda}$}, i.e., the probability for a market trade to happen with a duration time less than $\delta$ is
\begin{equation}
P(\delta) = \frac{1}{\lambda} \int_0^\delta e^{-t/\lambda} dt = 1 - e^{-\delta/\lambda}.
\label{cumprob}
\end{equation}

If a dark fill happens at a random time, the duration between the fill and the next lit market print is also a Poisson process with the same rate $\lambda(t)$. The reason for this is somewhat counter-intuitive (indeed, there is a ``waiting paradox'' named after it) and is due to the memory-less property of a Poisson process.

\sep[Estimating the wait surprise]
For a dark fill we can estimate the likelihood (surprise) of the duration between the fill and the next lit trade under the locally estimated Poisson intensity. Having observed $n$ lit market durations we need a probability density function for the next predicted print duration. This is done via a plug-in distribution, which we construct by plugging in an appropriate estimate of the intensity $\hat{\lambda}$ into the exponential density function
\begin{equation}
p(\delta_{n+1}|\hat{\lambda}) = \frac{1}{\hat{\lambda}} e^{-\delta_{n+1}/\hat{\lambda}}.
\end{equation}
In order to take into account the reliability of the local intensity estimate (which is influenced by how many lit trades prior to the dark fill are used to estimate the intensity, and the variability of durations) we can use a Bayesian approach. In this case, a Bayesian posterior distribution using an information-less objective prior $1/\lambda$ is 
\begin{equation}
\label{prob}
p(\delta_{n+1} | \delta_1, \delta_2, ..., \delta_n) = \frac{n^{n+1} \bar{\delta}^n}{(n \bar{\delta} + \delta_{n+1})^{n+1}},
\end{equation}
where $\bar{\delta}$ is the mean duration. In a frequentist's world, this expression is also a probability function obtained by maximising the joint likelihood function from which $\hat{\lambda}$ was eliminated~\cite{freq1,freq2,freq3}.


Analogously to Eq.~\ref{cumprob}, the probability of observing a duration $\delta_{n+1}$ having previously observed $n$ durations is obtained by integrating
\begin{eqnarray}
P( \delta_{n+1}) &=& \int_0^{\delta_{n+1}} p(\delta| \delta_1, \delta_2, ..., \delta_n) d\delta \nonumber \\
&=& 1 - \frac{\left( n \bar{\delta}\right)^n}{\left(n \bar{\delta} + \delta_{n+1}\right)^n}
\end{eqnarray}

In the context of our problem, $\delta_{n+1}$ is the duration between the dark fill and the next lit market print, while the previous $n$ durations $(\delta_1, \delta_2, ..., \delta_n)$ are lit market durations. 

More important for this application is one minus the probability $1-P(\delta_{n+1})$, which can be understood as the p-value of the test that the $\delta_{n+1}$ duration is consistent with the estimated local Poisson process, i.e., that duration $\delta_{n+1}$ was generated by the same Poisson process that generated the previous $n$ durations. As with any p-test, low values are considered as evidence against the hypothesis, in this case that the post dark fill duration was consistent with background market activity at the time. In other words, a dark fill quickly followed by lit trading activity and resulting with a p-value of, say, 0.01, we would expect to happen by chance in less than 1 out of 100 cases. It is more likely that the lit trades were triggered by the dark fill.


\begin{figure}[t]
\includegraphics[width=0.5\textwidth]{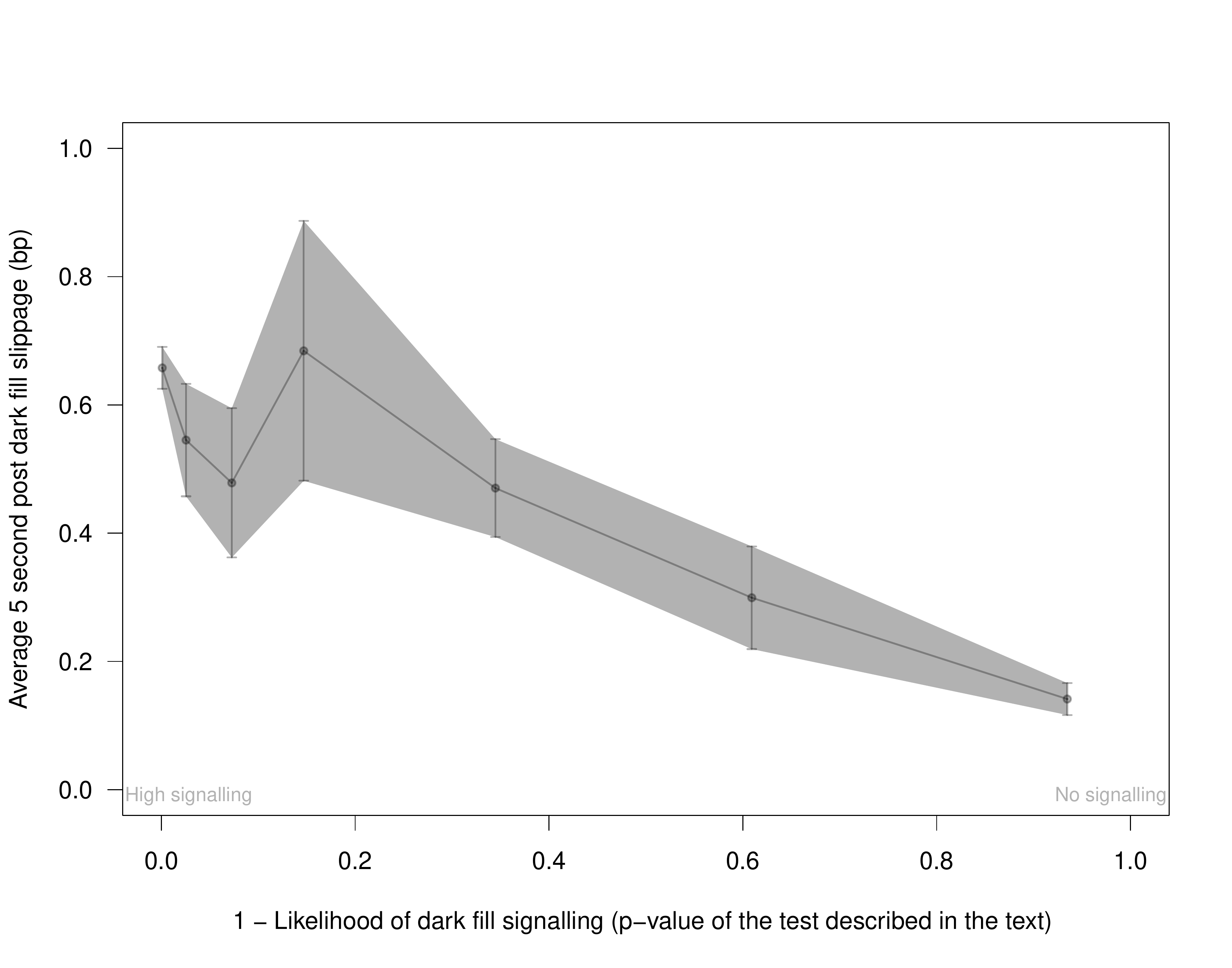}
\caption{
{\bf Dark fill slippage is significantly higher for fills quickly followed by lit prints.}
Chart shows the average dark fill slippage as a function of the likelihood of signalling as described in the text.
Slippage, measured as the 5 second post fill signed price move, increases with the "surprise factor" in the timing of the lit prints following the dark fill.
The surprise is measured by the p-value of the test described in the text and is equal to 1 minus the likelihood (probability) that the next lit trade following the dark fill has a duration consistent with previous lit market activity, i.e., that its timing is not unusual in the context of market activity at the time. We can see that the fills which have a high surprise factor, or high likelihood of signalling (values on the x-axis close to zero), have a significantly higher average slippage than fills with a low likelihood of signalling (x-axis values towards one).
\newline
{\tiny Data from start of April to end of October 2016. Dark fills executed by Barclays 
\emph{Hydra dark only} strategy in 100 London Stock Exchange (LSE) names selected by the most number of executed fills by Barclays. LSE market data used from Reuters.}
}
\label{slipVsPval}
\end{figure}

\sep[Empirical evidence]
Figure~\ref{slipVsPval} shows the empirical relation between the average dark fill slippage and signalling evidence as quantified by the p-value of the described test. On the y-axis, the average 5 second slippage is computed as a function of the bucketed p-value on the x-axis. As can be seen, x-axis values close to zero, corresponding to significant signalling evidence have a threefold higher post fill adverse price move than fills which carry no evidence of signalling. Slippage is computed by averaging the signed price return 5 seconds after in each bucket.

\sep[Optimally merging the evidence of individual fills]
To use this method in practice, however, we can not rely on a large number of fills required to compute the above averages. To be applicable in a reactive setting and inform the trader to dynamically tailor liquidity exposure, the methodology needs to discriminate liquidity with only a few fills done. Each fill with undesired liquidity is potentially costly, and we need to make the best use of the important evidence each fill brings.

Obviously, a single dark fill with surprising following lit print timing does not quite constitute reliable evidence of signalling, and most likely should not be reacted to. But once the trader receives a few fills from a venue, they should be able to reliably understand the quality of the liquidity trading on the venue at the moment, and decide whether they want to maintain or modify access to the venue.

To quantitatively do this, we need to efficiently merge the evidence collected by the $k$ previously traded fills in such a way that, even if each of the $k$ fills taken in isolation does not carry evidence of signalling, when considered together as a sequence of events, we can establish the overall likelihood of signalling.

To this end, we can make use of the Fisher's method~\cite{fisher} for repeated experiments, which consists of recognising that the log sum of uniform random variables (which p-values are) is distributed according to a Chi-squared distribution with $2k$ degrees of freedom
\begin{equation}
\label{fisher}
-2 \sum_{i=1}^k \log(p_i) \sim \chi_{2k}^2.
\end{equation}

Therefore for each group of $k$ dark fills, by computing the above sum, we can read off from Chi-squared the likelihood that the dark fills are leaking information and are detrimental to the order execution. In other words, each fill from, say a particular exchange, can be thought of as an experiment bringing in the valuable information about the liquidity available to the trader on the venue at the time. By using Fisher's method, they optimally use this information, allowing to quickly accumulate evidence of potential leakage and appropriately adjust their liquidity exposure.

\begin{figure}[t]
\includegraphics[width=0.5\textwidth]{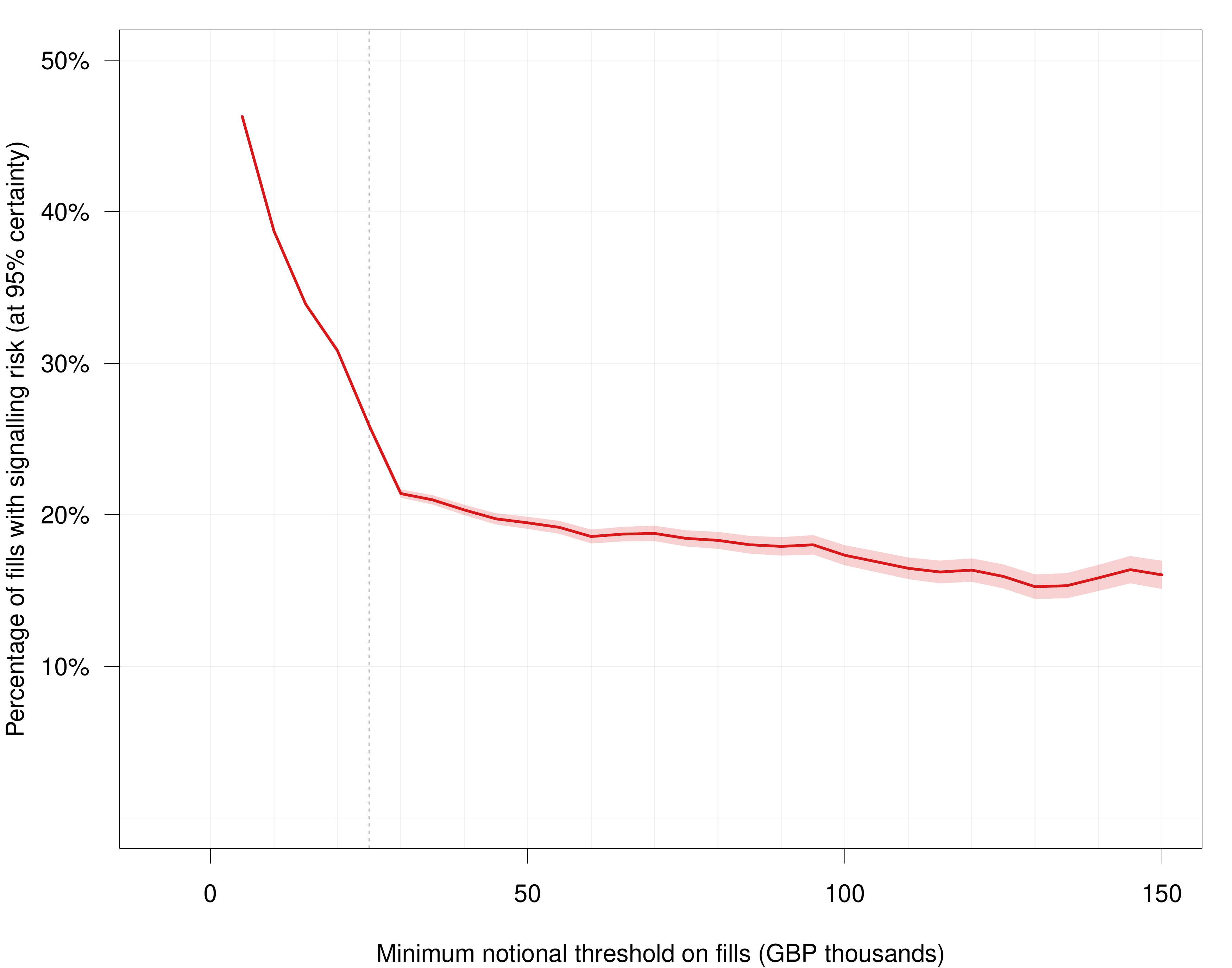}
\caption{
{\bf Requiring a minimum fill size on dark fills significantly reduces signalling risk -- however, it also reduces liquidity exposure.}
The y-axis displays the percentage of dark fills in our sample for which there is evidence of signalling to the lit markets. On the x-axis we show how these percentages might change if a minimum dark fill size is imposed. We can see that by limiting fills to be larger than \pounds 25,000 we substantially reduce the percentage of signalling dark fills. However, as emphasized in the text, doing this also significantly reduces the liquidity available to the trader. Signalling risk is estimated by measuring the time difference between the dark fill and the next print in the lit markets, and comparing this duration with the background lit market activity at the time. 
\newline
{\tiny Data from start of April to end of October 2016. Dark fills executed by Barclays \emph{Hydra dark only} strategy in 100 London Stock Exchange (LSE) names selected by the most number of executed fills by Barclays. LSE market data used from Reuters.}
}
\label{sizeFig}
\end{figure}

Once a trader observes a few fills from a venue which indicate there is information leakage, the trader may in the extreme avoid filling on the venue, but more likely may impose a minimum fill size for the  venue. That this is an effective strategy can be seen in figure~\ref{sizeFig} in which we show that dark fills restricted to large sizes indeed do show significantly less signalling and information leakage. Based on a sample of Barclays dark venue fills, the chart's y-axis shows the percentage of dark fills with signalling evidence that have a fill size greater than the x-axis GBP size threshold. Evidence of signalling is determined by the timing of lit prints following the dark fill and the test p-value being smaller than 0.05. 
Only about 20\% of dark fills larger than \pounds 30,000 are quickly followed by lit market prints, indicating signalling. In comparison, almost 50\% of dark fills when unrestricted in fill size are quickly followed by lit trades. Curiously, increasing minimum fill size beyond about \pounds 30,000 does not seem to reduce signalling. Its effect at that point is only to reduce the available liquidity a trader is interacting with.

\sep[Validating the approach on Barclays own dark fills]
Providing an estimate on the expected real world execution improvements of this methodology is difficult. However, a crude way to assess the statistical and practical power of the test may be done using Barclays algo execution data. The data used is from algos trading only in dark venues (\emph{Hydra dark only}) from start of April to end of October 2016. The analysed symbols were restricted to a set of 100 London Stock Exchange names selected by the most number of executed fills by Barclays.

We furthermore limit the analysis to algo executions requiring more than 15 fills to complete and use the arrival price slippage\footnote{Arrival price slippage is the order signed difference between the volume weighted average order price and the arrival mid price. We proxy the arrival mid price by the first fill of the order, which being a dark order is almost always the mid price.} as a comparison metric. The possible increased liquidity exposure this methodology allows can not be tested using historical data.

In order to be realistic, for each algo execution we compute the likelihood of signalling (using Fisher's method) \emph{using only the first 5 fills of an order execution}, irrespectively of the total number of fills an order eventually generated. The reason for using only a few initial fills is that in practice we would normally change the algo behaviour upon detecting potential signalling. 
We find that algo executions with substantial signalling risk -- determined by the p-value arising from Eq.~\ref{fisher} at the 5\% level -- have roughly 30\% higher absolute slippage values than executions for which we do not find evidence of signalling risk.

\section{Conclusion}
\sep[Nature of dark liquidity]
Dark liquidity in equities markets today is not fully dark, allowing for fills with negligible market impact. It may better be described by shades of gray, with different signalling risks associated with different types of liquidity. Clips of large algo executions will be interspersed with information seeking orders, making it difficult to distinguish between the two.

Furthermore, liquidity principally is dynamic in nature, its composition constantly changing with traders entering and exiting the market, working their orders. 

\sep[Pitfalls of common methods]
Simple static methods, such as requiring a minimum fill size, while effective at reducing signalling, also result in a significantly reduced access to liquidity as interaction with algo orders is limited. 

Dynamical methods based on estimating slippage are not a viable option either. Too many fills are required to reliably detect slippage for it to be usable in a setting where reactivity is required. As a consequence, methods that monitor slippage are more likely reacting to random market fluctuations than signalling.

On the other hand, slippage, if at all detectable, is not a sufficient piece of information for a trader to react in a meaningful way. If the slippage is \emph{not caused} by the trader's fills, the reaction to scale trading back is quite the opposite to what a meaningful response would be. Observing slippage that is not caused by own orders may be an indication of competing traders for the available liquidity and trading should be scaled up, not down.

\sep[Fisher's method and its benefits]
The method we describe in this paper colours the dark liquidity by dynamically detecting evidence of causality between dark and lit fills and provides reactive information for a trader to adjust their trading. It is based on monitoring unusual timing synchronicity of dark and lit fills, which for a fully dark venue should not be present. 

Using this method, a trader does not need to restrict access to liquidity at the start of an execution; on the contrary, they can target dark liquidity in all sizes and venues. As fills are generated, more and more evidence is accumulated, coloring the gray liquidity, and allowing the trader to reactively and dynamically restrict access to venues and sizes which, \emph{at the time}, they do not wish to interact with. Depending on the amount of evidence each fill brings in, different number of fills will be required to color the liquidity -- in the extreme cases, two to three fills may be sufficient information to pull back from accessing certain liquidity, protecting the trader.

This method can easily be extended to include other relevant lit market events (e.g. quote changes) and analyse their relation to dark fill timings. Ultimately the method applies naturally to market trades in multiple venues (not only dark) provided it is possible to exclude own fills from trade print feeds.

\begin{centering}
 \vspace {0.2cm}
\hrule
\vspace {0.8cm}
\sf \tiny 
{\sc Disclaimer} \\
 \vspace {0.2cm}
Barclays is a full service investment bank. In the normal course of offering investment banking products and services to clients. Barclays may act in several capacities (including issuer, market maker, underwriter, distributor, index sponsor, swap counterparty and calculation agent) simultaneously with respect to a product, giving rise to potential conflicts of interest which may impact the performance of a product. This document is from a Barclays Trading and/or Distribution desk and is not a product of the Barclays Research department.  Any views expressed may differ from those of Barclays Research. Barclays, its affiliates and associated personnel may at any time acquire, hold or dispose of long or short positions (including hedging and trading positions) which may impact the performance of a product.  \vspace {0.1cm}

This document is provided for information purposes only and it is subject to change. It is indicative only and is not binding.
Barclays is not offering to sell or seeking offers to buy any product or enter into any transaction. Any transaction requires Barclays€™ subsequent formal agreement which will be subject to internal approvals and binding transaction documents.  Without limitation to the foregoing, any transaction may also be subject to review by Barclays against its published Tax Principles. Barclays is not responsible for the use made of this document other than the purpose for which it is intended, except to the extent this would be prohibited by law or regulation.   \vspace {0.1cm}

Obtain independent professional advice before investing or transacting. Barclays is not an advisor and will not provide any advice relating to a product. Before making an investment decision, investors should ensure they have sufficient information to ascertain the legal, financial, tax and regulatory consequences of an investment to enable them to make an informed investment decision. Barclays is not responsible for information stated to be obtained or derived from third party sources or statistical services. Any past or simulated past performance (including back-testing) contained herein is no indication as to future performance. \vspace {0.1cm}

All opinions and estimates are given as of the date hereof and are subject to change. Barclays is not obliged to inform investors of any change to such opinions or estimates. This document is being directed at persons who are professionals and is not intended for retail customer use. For important regional disclosures you must read, click on the link relevant to your region.  Please contact your Barclays representative if you are unable to access. \vspace {0.1cm}

EMEA https://www.home.barclays/disclosures/important-emea-disclosures.html. 
APAC https://www.home.barclays/disclosures/important-apac-disclosures.html. 
U.S. https://www.home.barclays/disclosures/important-us-disclosures.html. 
This document is confidential and no part of it may be reproduced, distributed or transmitted without the prior written permission of Barclays. Barclays offers premier investment banking products and services to its clients through Barclays Bank PLC.  Barclays Bank PLC is authorised by the Prudential Regulation Authority and regulated by the Financial Conduct Authority and the Prudential Regulation Authority and is a member of the London Stock Exchange. Barclays Bank PLC is registered in England No. 1026167 with its registered office at 1 Churchill Place, London E14 5HP.  Barclays Capital Inc. is a US registered broker/dealer affiliate of Barclays Bank PLC and a member of SIPC, FINRA and NFA.\vspace {0.1cm}
  
© Copyright Barclays Bank PLC, 2017 (all rights reserved).  

\end{centering}


\begin{thebibliography}{1}
\bibitem{freq1} Lawless, J.F., Fredette, M.,"Frequentist predictions intervals and predictive distributions", Biometrika (2005), Vol 92, Issue 3.

\bibitem{freq2} Bjornstad, J.F. (1990). "Predictive Likelihood: A Review", Statist. Sci. 5 (2).

\bibitem{freq3} D. F. Schmidt and E. Makalic, "Universal Models for the Exponential Distribution", IEEE Transactions on Information Theory (2009), Volume 55, Number 7.

\bibitem{fisher} Fisher, R.A. (1925). Statistical Methods for Research Workers. Oliver and Boyd (Edinburgh).

\bibitem{farmer} Farmer, J.D. "Market Force, Ecology, and Evolution", Industrial and Corporate Change 11(5) (2002): 895-953.

\end{thebibliography}
\end{document}